\documentclass[aps,prb,twocolumn,showpacs,floats,preprintnumbers,amsmath,amssymb]{revtex4}

\usepackage{graphicx}% Include figure files
\usepackage{dcolumn}% Align table columns on decimal point
\usepackage{bm}% bold math

\begin{document}

\preprint{(submitted to Phys. Rev. B)}

\title{Stabilizing a Molecular Switch at Solid Surfaces:\\
A Density-Functional Theory Study of Azobenzene at Cu(111), Ag(111), and Au(111)}

\author{Erik McNellis}
\author{J\"org Meyer}
\author{Abbas Dehghan Baghi}
\altaffiliation{Present address: I.C.T.P Affiliated Center, Physics Department, Isfahan University of Technology, Isfahan-Iran.}
\author{Karsten Reuter}
\affiliation{Fritz-Haber-Institut der MPG, Faradayweg 4-6, D-14195 Berlin (Germany)}

\received{05.03.09}

\begin{abstract}
We present a density-functional theory trend study addressing the binding of the trans-cis conformational switch azobenzene (C$_6$H$_5$-N=N-C$_6$H$_5$) at three coinage metal surfaces. From the reported detailed energetic, geometric, and electronic structure data we conclude that the governing factor for the molecule-surface interaction is a competition between covalent bonding of the central azo (-N=N-) bridge on the one hand and the surface interaction of the two closed-shell phenyl (-C$_6$H$_5$) rings on the other. With respect to this factor the cis conformer exhibits a more favorable gas-phase geometric structure and is thus more stabilized at the studied surfaces. With the overall binding still rather weak the relative stability of the two isomers is thereby reduced at Ag(111) and Au(111). This is significantly different at Cu(111), where the cis bonding is strong enough to even reverse the gas-phase energetic order at the level of the employed semi-local electronic exchange and correlation (xc) functional. While this actual reversal may well be affected by the deficiencies due to the approximate xc treatment, we critically discuss that the rationalization of the general effect of the surface on the meta-stable molecular states is quite robust. This should equally hold for the presented analysis of recent tip-manipulation and photo-excitation isomerization experiments from the view point of the derived bonding mechanism.
\end{abstract}


\maketitle

\section{Introduction}

In view of the rapidly advancing miniaturization in microelectronics and sensing, molecules are envisioned as fundamental building blocks in a future ``molecular nanotechnology''. Since controlled switching between defined states is a crucial basis component for storage and logic, molecules offering this functionality (e.g. through externally induced changes between conformational isomers) attain a central importance. Considering contacting and defined integration into a larger framework, it is more precisely the function of the molecule when stabilized at a solid surface that is of key interest. While a large variety of molecules can be controllably switched in gas-phase or solution, still little is known about their function in such an adsorbed state. Suppression of the switching capability e.g. due to steric hindrance is a possible scenario, but completely new properties under the influence of the solid surface are equally conceivable. The atomic-scale understanding necessary for a technological exploitation of such effects builds on a detailed structural and electronic characterization of the adsorbed molecular switch, as well as its response to external stimuli like fields, forces or external currents.\cite{applphysa}

As a prototypical conformational switch of modest complexity azobenzene (C$_6$H$_5$-N=N-C$_6$H$_5$) has been on the research agenda for many years.\cite{rau90} In solution it exhibits a reversible photo-isomerization between an energetically more stable planar trans geometry and a three-dimensional cis configuration, in which the planes of the two phenyl-rings are tilted with respect to each other. A common theme in existing experimental studies addressing the function in surface mounted geometries is that a too strong substrate interaction is suspected to cause adverse effects on the switching efficiency. This has either led to the use of ligands like alkanethiol chains to decouple the azobenzene moiety in vertical geometries \cite{pace07,schmidt08} or to a focus on unreactive materials like close-packed surfaces of coinage metals and there in particular on Au(111). On the latter surface bare azobenzene could be successfully isomerized using tip-manipulation techniques \cite{choi06}, but even in this alleged weak physisorption limit switching with light could not be achieved \cite{comstock07}. Reasoned as an ultrafast quenching of the electronic excitation azobenzene-derivatives that are further lifted up from the surface appear as a viable alternative and reversible photomechanical switching has indeed been reported for the functionalized molecule tetra-tert-butyl azobenzene (TBA) \cite{comstock07,hagen07}. The intricate role played even then by the metallic substrate is nevertheless exemplified by the fact that exactly for the same azobenzene-derivative, TBA, light-induced switching could not be achieved at Ag(111).\cite{tegeder07}

As already stated, the characterization of the stable (or meta-stable, long-lived) molecular states at the surface is a necessary prerequisite to a detailed understanding of these experimental findings and of the actual excitation mechanism. While already less ambitious than a comprehensive first-principles treatment of the entire switching process, corresponding static electronic structure theory calculations still pose a considerable challenge. The large system sizes resulting from the sheer extension of the azobenzene molecule and the periodic supercell geometries dictated by the metallic bandstructure, can at present only be tackled by density-functional theory (DFT) with local- or semi-local exchange and correlation (xc) functionals. While this is the state of the art and the lowest level of theory at which one can still at least hope for the aspired quantitative and predictive-quality modeling, it is clear from the start that the specificities of the molecule-surface binding directly challenge two well-known deficiencies of the named functionals: the spurious self-interaction and the lack of long-range dispersive (van der Waals) attraction. With several frontier orbitals of different symmetry ({\em vide infra}) presumably involved in the bonding, the detrimental effect of self-interaction blurred orbital energies on fundamental quantities like the preferred adsorption site has already been demonstrated for much simpler adsorbates \cite{kresse03,mason04,hu07,stroppa08,sharifzadeh08}. Similarly well known is the important role played by long-range van der Waals interactions when aromatic $\pi$-like orbitals participate in the molecule-surface interaction \cite{hausschild05,ortmann05,atodiresei08}.

The present trend study addressing the binding of azobenzene at the three coinage metal surfaces, Cu(111), Ag(111) and Au(111) within large-scale DFT calculations correspondingly has a twofold focus. On the one hand, we critically discuss the obtained detailed energetic, geometric, and electronic structure data in the context of the sketched limitations of the employed semi-local xc functional, namely the generalized gradient approximation (GGA) functional due to Perdew, Burke and Ernzerhof (PBE) \cite{perdew96}. With respect to the lack of long-range dispersive interactions this includes the main conclusions from an assessment of this issue on the level of semi-empirical van der Waals correction schemes \cite{grimme06,ortmann06,tkatchen09}, the details of which will be published in a consecutive paper \cite{mcnellis09}. On the other hand, we carve out some governing factors in the molecule-surface interaction that appear quite robust with respect to the employed approximate xc treatment. This concerns notably a rationalization of the obtained energetic ordering of the molecular states at the surface, and an analysis of the recent tip-manipulation and photo-excitation experiments from the view point of the derived bonding mechanism. As such, the reported data also provides an important reference for future studies, in which we will investigate the differences induced by the engineered functionalization of this molecule to improve the switching efficiency.

\section{Theory}

The DFT GGA-PBE calculations were carried out within the plane-wave implementation and using the standard library ultrasoft pseudopotentials \cite{vanderbilt90} as provided by the CASTEP code \cite{clark05}. Gas phase molecular geometries and properties were determined by optimizing isolated molecules in a rectangular (30\,{\AA} $\times$ 40\,{\AA} $\times$ 30\,{\AA}) supercell, with $\Gamma$-point sampling and a plane-wave kinetic energy cutoff of 800\,eV. These supercell molecular calculations were complemented by computations of truly isolated molecules using the NWChem package \cite{NWChem} and the def-TZVP \cite{NWChembasis} basis set. While the geometries, orbital energies and relative stabilities of neutral azobenzene obtained with both codes for GGA-PBE are identical on the accuracy level discussed here, we additionally used the molecular NWChem code for reference calculations using the hybrid B3LYP functional \cite{becke93}, as well as for the calculation of ionized molecular properties, cf. Table \ref{table4} below. 

\begin{figure}
\centering
\includegraphics[width=3.8cm]{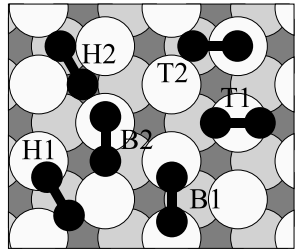}
\caption{Schematic top view indicating the initial lateral position of the two azo-bridge N atoms in the six geometry optimizations carried out to determine the most stable adsorption site at each surface (see text for nomenclature and details). N atoms, first- and second layer metal atoms are shown as black small circles, large light gray and large dark gray circles, respectively.}
\label{fig1}
\end{figure}

The molecule-surface system was modeled in supercell geometries, employing the GGA-PBE optimized lattice constants ($a_{\rm o}({\rm Cu}) = 3.63$\,{\AA}, $a_{\rm o}({\rm Ag}) = 4.14$\,{\AA}, $a_{\rm o}({\rm Au}) = 4.19$\,{\AA}) and, depending on the registry of the trans isomer with the substrate, large $(6 \times 3)$ or $(6 \times 4)$ (111) surface unit-cells to decouple the adsorbed molecule from its periodic images. No possibility for reconstruction was considered for both clean and adsorbate-covered surfaces. The search for the most stable adsorption site at concomitant optimized adsorption geometries was performed using three layer slabs with adsorption on one side and allowing full relaxation of the topmost layer. Structure optimization relied on the BFGS algorithm \cite{bfgs} as implemented in CASTEP, and the default convergence threshold on the maximum absolute ionic force component of 0.05\,eV/{\AA}. The plane-wave kinetic energy cutoff was 350\,eV, the vacuum region exceeded 13\,{\AA}, and reciprocal space integration was done using $(2 \times 4 \times 1)$ Monkhorst-Pack grids \cite{monkhorst99}. Centrally targeted in these calculations was the relative energetic stability, which we found to be converged to $\pm 20$\,meV at these computational settings. In order to determine the preferred adsorption site, both azobenzene isomers were optimized in six different adsorption geometries on all three surfaces as sketched in Fig. \ref{fig1}: The molecule was centered on the bridge (B), top (T) and hollow (H) surface sites, and shifted such that either the azo-bridge ($-\!{\rm N}\!=\!{\rm N}\!-$) center of mass (designated as 1) or one of the azo-bridge N atoms (designated as 2) aligned with the adsorption site. In most optimizations, the molecule relaxed towards the 1:1 metal-N atom coordinated B1 geometry (shown centered at the bottom of Fig. \ref{fig1}), which at all three surfaces and for both isomers was also the energetically most favorable among those found.

For the thus defined most stable geometry the absolute adsorption energy, work function and projected density of states (PDOS) were determined at refined computational settings and a $(6 \times 3)$ (111) surface unit-cell. To also avoid a finite dipole across the vacuum region we switched to thicker, inversion-symmetric seven layer slabs with adsorption on both sides. The vacuum was increased to 22\,{\AA} and the \textbf{k}-point density doubled. The adsorption energy is defined as
\begin{equation}
\label{eq1}
E_{\rm ads} \;=\; \frac{1}{2} \left[ E_{\rm azo@(111)} - E_{(111)} - E_{\rm azo(gas)} \right] \quad ,
\end{equation}
where $E_{\rm azo@(111)}$ is the total energy of the relaxed azobenzene-surface system, $E_{(111)}$ the total energy of the clean slab, $E_{\rm azo(gas)}$ the total energy of the corresponding relaxed gas-phase isomer (all three computed at the same plane-wave cutoff), and the factor $1/2$ accounts for the fact that adsorption is at both sides of the slab. The adsorption energy of either cis or trans isomer at the surface is thus measured relative to its stability in the gas-phase, and a negative sign indicates that adsorption is exothermic. Systematic convergence tests indicate that this central quantity is converged to within $\pm 30$\,meV at the chosen settings. 

The work function of a particular surface structure is determined as the difference of the Fermi level and the value of the electrostatic potential at the center of the vacuum between two consecutive slabs, and we also verified that this quantity is converged to within $\pm 30$\,meV at the chosen settings. As a technical point, we note that this required increasing the size of the grid on which the charge density is represented from the 1.8 $\mathbf{k}_{\rm cut}$ otherwise used, to the maximum of 2.0 $\mathbf{k}_{\rm cut}$. Additionally, we had to lower the SCF convergence energy threshold from otherwise $10^{-7}$\,eV/atom to $5 \cdot 10^{-9}$\,eV/atom. 

The molecular PDOS\cite{lorente00} $\rho_{a}(E)$ is defined by projections of the Kohn-Sham (KS) states of the (adsorbed) molecule-surface system  $\left| \phi_{b,k}\right\rangle$ with KS Eigenvalues $\epsilon_{b,k}$ onto a certain KS orbital $\left| \varphi_{a} \right\rangle$ of the isolated molecule in the adsorption geometry
\begin{equation}
  \label{eq_pdos}
  \rho_{a}(E) \;=\; \sum_{b,k} w_{k} \left|\left\langle \varphi_{a}\mid\phi_{b,k}\right\rangle \right|^{2}\delta(E-\epsilon_{b,k}) \quad .
\end{equation}
The summation over band indices $b$ and \textbf{k}-point indices $k$ is weighted by the \textbf{k}-point-weights $w_{k}$ so that an integration of $\rho_{a}(E)$ over $E$ gives $\left|\left\langle \varphi_{a} \mid \varphi_{a} \right\rangle \right|^{2}$ if sufficiently many bands of the molecule-surface system are considered. Conveniently, we used molecular KS orbitals $\left| \varphi_{a,k} \right\rangle$ which were calculated in the same supercell and with the same energy cut-off and \textbf{k}-point grid as the molecule-surface system to evaluate Eq.~(\ref{eq_pdos}). We verified that the artificial dispersion of these molecular KS orbitals is indeed negligible within the desired accuracy of our computational setup. In addition, Eq.~(\ref{eq_pdos}) was evaluated using the scalar product induced by the overlap matrix S due to the ultrasoft pseudopotentials \cite{vanderbilt90}. In the plots below, the $\delta$-functions have finally been convoluted with a Gaussian function of 0.2\,eV full width at half maximum for better visualization.

\section{Results and discussion}

\subsection{Gas-phase azobenzene}

\begin{table}
\caption{\label{table1}
Selected structural parameters of gas-phase azobenzene (AB) as defined in the text and in Fig. \ref{fig3} below. Compared are the optimized values from the GGA-PBE calculations against higher-level theory (B3LYP (this work) and MP2/cc-pVTZ \cite{fliegl03}) and experiment\cite{mostad71,bouwstra83}.}
\begin{ruledtabular}
\begin{tabular}{l|ccc|ccc}
         & \multicolumn{3}{c|}{trans AB} & \multicolumn{3}{c}{cis AB} \\
         & $d_{\rm NN}$  & $\omega$  & $\alpha$  & $d_{\rm NN}$  & $\omega$  & $\alpha$ \\
         & ({\AA})       & (deg)     & (deg)     & ({\AA})       & (deg)     & (deg)    \\[0.1cm] \hline
GGA-PBE  & 1.30          & 0         & 115       & 1.28          & 12        & 124      \\
B3LYP    & 1.25          & 0         & 115       & 1.24          & 9         & 124      \\
MP2      & 1.27          & 0         & 114       & 1.26          & 7         & 121      \\[0.2cm]
Exp.     & 1.25          & 0         & 114       & 1.25          & 8         & 122      \\
\end{tabular}
\end{ruledtabular}
\end{table}

We begin our analysis with a short compilation of data for gas-phase azobenzene that is of relevance for the ensuing study of the surface mounted switch. With the molecular structure well summarized as two phenyl rings connected by an azo ($-\!{\rm N}\!=\!{\rm N}\!-$) bridge, central geometric parameters are the azo-bridge bond length $d_{\rm NN}$, as well as the two angles that describe the orientation of each symmetry-equivalent phenyl-ring to the azo-bridge and of the phenyl-rings to each other, inversion angle $\alpha$ and dihedral angle $\omega$, respectively. Figure \ref{fig3} below illustrates these parameters in the context of the surface adsorbed geometry, while Table \ref{table1} compares their optimized gas-phase values obtained from the GGA-PBE calculations with higher-level theory and experiment. With the possible exception of a slightly elongated azo-bridge bond length it is fair to say that with respect to the geometric structure the employed GGA functional yields overall results that are as good as that of the hybrid B3LYP functional or the Hartree-Fock based M{\o}ller-Plesset perturbation theory (MP2). This favorable performance also carries over to the relative energetic stability of the two isomers. Here, we find the trans form to be more stable by 0.57\,eV than the cis form, in good agreement to the values of 0.68\,eV and $\sim 0.6$\,eV from the B3LYP calculations and from experiment \cite{schulze77}, respectively. 

\begin{figure}
\centering
\includegraphics[width=7.8cm]{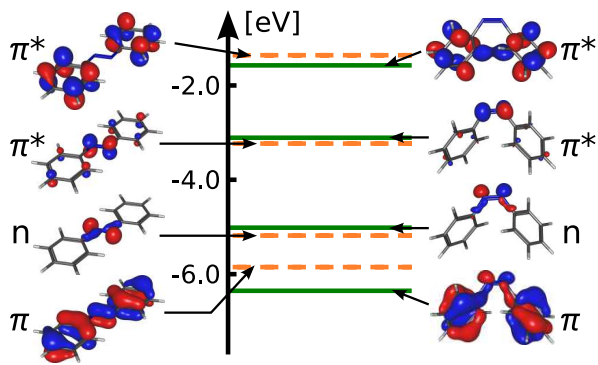}
\caption{(Color online) GGA-PBE Kohn-Sham eigenvalues with respect to the vacuum level of the four frontier orbitals HOMO-1 ($\pi$), HOMO ($n$), LUMO ($\pi^*$) and LUMO+1 ($\pi^*$) of gas-phase trans (solid lines) and cis (dashed lines) azobenzene. Additionally shown are corresponding orbital representations superimposed on the skeleton geometric structure (left: trans, right: cis).}
\label{fig2}
\end{figure}

As anticipated the description of the molecular frontier orbitals is much more problematic. Figure \ref{fig2} shows representations of those KS orbitals that are of particular interest both with respect to the bonding mechanism to the surface, and with respect to possible isomerization pathways. These range from the second highest occupied molecular orbital (HOMO-1), over HOMO and lowest unoccupied molecular orbital (LUMO) to LUMO+1. Compared to e.g. the much less self-interaction hampered B3LYP functional the GGA-PBE functional still yields the correct molecular character of these orbitals. In the higher-symmetry trans state the HOMO-1 and LUMO thus contain large contributions of the $\pi$ and $\pi^*$ orbital of the $-\!{\rm N}\!\!=\!\!{\rm N}\!-$ moiety, respectively. The HOMO consists predominantly of the N lone pairs and the LUMO+1 is centered on the $\pi^*$ system of the phenyl rings. This character of the four frontier orbitals is also largely retained in the cis form, which is why they are often still classified as $\pi$, $n$, or $\pi^*$ despite the lack of symmetry of this isomer. While the ordering of these frontier orbitals is thus still identical in B3LYP and GGA-PBE calculations, large differences exist in their actual energetic position. This is dramatically illustrated by the computed HOMO-LUMO gap, which in GGA-PBE comes out at respectively 1.95\,eV and 1.91\,eV for trans and cis azobenzene, and which is therewith about 2\,eV smaller than at the B3LYP level (trans: 3.92\,eV, cis: 3.76\,eV). The corresponding self-interaction driven underestimation of the HOMO-LUMO gap is typical for a GGA functional \cite{cohen08} and the crucial question is how much of this error is retained in the adsorbed state.

\subsection{Adsorption geometry and energetics}

\begin{figure}
\centering
\includegraphics[width=7.8cm]{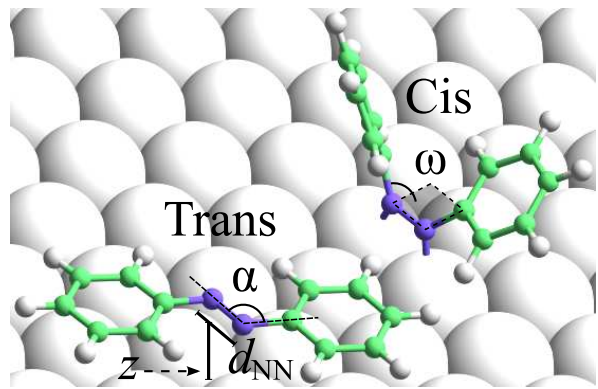}
\caption{(Color online) Perspective view of adsorbed trans and cis azobenzene in the most stable B1 adsorption site, cf. Fig. \ref{fig1}, illustrating the main parameters defining the adsorption geometry (see text): The azo-bridge bond length $d_{\rm NN}$ and the vertical height $z$ of the azo-bridge above the surface plane defined by the average height of the first-layer atoms. Furthermore the azo-bridge to phenyl-ring bond angles $\alpha$ and the azo-bridge dihedral angle $\omega$. The latter is defined as the smallest angle formed between the $-$C$-$N$-$ bond to one phenyl-ring and the plane spanned by the $-\!{\rm N}\!\!=\!\!{\rm N}\!-$ bridge and the $-$N$-$C$-$ bond to the other phenyl-ring.}
\label{fig3}
\end{figure}

As already stated in Section II the most favorable adsorption site obtained at all three coinage metal surfaces corresponds to the $-\!{\rm N}\!\!=\!\!{\rm N}\!-$ moiety bridging between two surface atoms as sketched in the B1 geometry of Fig. \ref{fig1}. While in this adsorbed state the azo-bridge could in principle be vertically tilted with a concomitant symmetry breaking of the molecular structure as expressed by two different inversion angles $\alpha$ of the two phenyl-rings, we find that this is not the case. At all three surfaces, the azo-bridge lies flat at a height $z$ above the surface, and the two phenyl-rings show identical inversion angles that are furthermore essentially unchanged compared to the gas-phase value, i.e. 115$^\circ$ and 124$^\circ$ for trans and cis azobenzene respectively, cf. Table \ref{table1}. Figure \ref{fig3} illustrates the resulting adsorption geometry and the main structural parameters discussed already in the context of the gas-phase molecule. 

\begin{table}
\caption{\label{table2}
Structural parameters as defined in Fig. \ref{fig3} for trans and cis azobenzene (AB) in the relaxed most stable adsorption geometry at the three coinage metal surfaces and compared to the optimized gas-phase structure.}
\begin{ruledtabular}
\begin{tabular}{l|ccc|ccc}
         & \multicolumn{3}{c|}{trans AB} & \multicolumn{3}{c}{cis AB} \\
         & $z$    & $d_{\rm NN}$  & $\omega$  & $z$ &  $d_{\rm NN}$  & $\omega$ \\
         & ({\AA}) & ({\AA})       & (deg)     & ({\AA})  & ({\AA})       & (deg)      \\[0.1cm] \hline
gas-phase & $-$  & 1.30 & 0  & $-$  & 1.28 & 12 \\[0.1cm]
Cu(111)   & 1.98 & 1.40 & 39 & 1.93 & 1.35 & 33 \\
Ag(111)   & 3.64 & 1.30 & 0  & 2.27 & 1.32 & 23 \\
Au(111)   & 3.50 & 1.30 & 0  & 2.31 & 1.29 & 18 \\
\end{tabular}
\end{ruledtabular}
\end{table}

If we start the detailed discussion with trans azobenzene at Au(111) and Ag(111), one can summarize the calculated geometry values listed in Table \ref{table2} in that the molecule remains essentially flat and floats at a rather large height $> 3.5$\,{\AA} above the surface. Essentially flat in this context means that the dihedral angle $\omega$ remains in both cases zero as in the gas-phase. Due to some torsional twisting of the phenyl-rings, the variation of $z$-heights exhibited by the individual atoms within the molecule nevertheless amounts to $\Delta z$ = 0.34\,{\AA} and 0.37\,{\AA} on Au(111) and Ag(111), respectively. While this is relatively small, it could nevertheless be an important feature considering that recent conformational dynamics calculations  \cite{henningsen07} for a free azobenzene derivative found that a non-planar geometry is
a necessary prerequisite to efficiently couple electronic excitations with intramolecular rotations.

\begin{table}
\caption{\label{table3}
Adsorption energies $E_{\rm ads}$ as defined in Eq. (\ref{eq1}) for trans and cis azobenzene at the three coinage metal surfaces, as well as the relative energetic stability of the two adsorbed isomers $\Delta E$ with $\Delta E > 0$ indicating a higher stability of the trans isomer. All values in eV.}
\begin{ruledtabular}
\begin{tabular}{l|ccc}
        & $E_{\rm ads}$(trans) & $E_{\rm ads}$(cis) & $\Delta E$ \\[0.1cm] \hline
Cu(111) & $-$0.27 & $-$1.08 & $-$0.24\\
Ag(111) & $-$0.11 & $-$0.42 & 0.26\\
Au(111) & $-$0.12 & $-$0.27 & 0.42\\
gas-phase& $-$  & $-$   & 0.57\\
\end{tabular}
\end{ruledtabular}
\end{table}

The comparatively weak bonding suggested by the large $z$-height is indeed confirmed by the actually calculated exothermic, but small adsorption energy of about -0.1\,eV at both surfaces, cf. Table \ref{table3}, and is furthermore consistent with the virtually unchanged azo-bridge bond length of 1.30\,{\AA} in the adsorption geometry, cf. Table \ref{table2}. The correspondingly deduced weak physisorptive bonding scenario is, however, particularly prone to the prominent lack of dispersive interactions in the employed GGA-PBE functional as discussed in the introduction. Without a corresponding attractive contribution, the $z$-height of the molecule is exclusively governed by the steep Pauli repulsion wall encountered by the phenyl-rings, preventing a closer approach and therewith stronger bonding of the molecule to the surface. Without further qualification of such interactions we would therefore refrain from putting too much emphasis on the actually calculated value of the $z$-height and the adsorption energy. In fact, our investigation addressing this issue at least on the level of semi-empirical van der Waals correction schemes indicates that the dispersive attraction induced lowering of the $z$-height could amount to up to $\sim 0.7$\,{\AA}, with a concomitant significant increase in the adsorption energy \cite{mcnellis09}.

Nevertheless, even at such a reduced height, the planar trans geometry fixes the azo-bridge atoms at a distance to the surface that is outside the range that would enable the formation of stronger covalent bonds to the top-layer metal atoms, and is thereby the main limiting factor in the molecule-surface interaction. This problem does not exist for the cis isomer, where due to the bent molecular geometry the azo-bridge can come much closer to the surface with the phenyl-rings pointing away from the surface and thus at heights that are not yet in conflict with the repulsive wall, cf. Fig. \ref{fig3}. This picture is nicely reflected by the computed much smaller $z$-heights slightly in excess of 2\,{\AA}, cf. Table \ref{table2}, that bring the azo-bridge atoms into bond distances to the directly coordinated surface metal atoms that are highly reminiscent of corresponding distances in transition metal complexes \cite{paramonov03}. As a consequence the actual intramolecular $-\!{\rm N}\!\!=\!\!{\rm N}\!-$ bond gets activated as reflected by the computed slightly elongated bond length compiled in Table \ref{table2}. In detail, the resulting optimum $z$-heights still seem to require some additional bending of the molecule to lift the phenyl-rings even further away from the surface as expressed by the somewhat increased dihedral angle $\omega$ for cis azobenzene at both surfaces, cf. Table \ref{table2}. The cost for this bending partly compensates the gain from the formed azo-bridge surface bonds, but as apparent from Table \ref{table3} the net resulting adsorption energy for the cis isomer is still larger than in the trans case, and correspondingly, we arrive at a decrease of the relative energetic stability of the two isomers. 

Nevertheless, at both Au(111) and Ag(111) the trans form is still the more stable one, consistent with the conclusion from several STM studies which interpreted the measured ``dumbbell'' appearance of the thermally more stable adsorbed molecular configuration in terms of the trans isomer \cite{choi06,comstock07,comstock05}. While DFT calculations based on a local-density approximation (LDA) \cite{ceperley80} xc functional supporting one of these STM investigations \cite{choi06} also found the trans isomer to be more stable, it is important to note that their reported bond strength to the surface is qualitatively different to our findings. For both isomers at Au(111) strong chemisorptive binding of more than 1.5\,eV was computed, with the most favorable cis adsorption geometry in fact featuring one phenyl-ring parallel to the surface and one phenyl-ring standing upright.\cite{choi06} Suspecting this difference to predominantly arise out of the use of the different, yet equally standard xc functional, this underscores the initially made statements concerning the sensitivity of the description of the molecule-surface interaction to the approximate treatment of electronic exchange and correlation in present-day functionals available for huge supercell calculations. When accounting for the lacking dispersive attraction in the here employed GGA-PBE functional with semi-empirical van der Waals schemes, we obtain a maximum correction of the adsorption energy of the order of 1\,eV \cite{mcnellis09}. While this brings the absolute value closer to the reported LDA result, the one phenyl-ring upright cis geometry is never found to be stable, let alone that it is clear that the LDA gives the bond strength for the wrong reasons. In this respect, we are confident that the approach of dispersion corrected GGA-PBE energetics employed in this study provides a much more adequate description of the azobenzene bonding to coinage metal surfaces, and we will critically discuss this point in more detail in our consecutive publication \cite{mcnellis09}.

Moving over to Cu(111) as the remaining of the three surfaces we find a much stronger interaction even at the GGA-PBE level. In contrast to the situation at Ag(111) and Au(111), both the trans and cis form of azobenzene adsorbs at a vertical $z$-height that brings the azo-bridge atoms to typical covalent N-Cu bond distances known from transition metal complexes \cite{szalda76}.
While due to the smaller Cu radius the Pauli repulsion for the phenyl-rings sets in at a closer distance \cite{mcnellis09}, such low $z$-heights are nevertheless incompatible with a fully planar trans adsorption geometry. As shown in Table \ref{table2} this leads to a pronounced buckling of the molecule with a final dihedral angle $\omega$ that is almost equal to the one of the adsorbed cis isomer, with even the latter angle being significantly increased with respect to its gas-phase value. The cost of this severe distortion of trans azobenzene again directly carries over to the final adsorption energy. While the strong surface-interaction indicated by the much elongated -N=N- moiety is directly visible in the strong $E_{\rm ads}$ of the cis isomer, the adsorption energy in the trans case is correspondingly much smaller despite an even more activated azo bond length. 

At Cu(111) this difference in the net adsorption energy finally leads to a reversal of the relative stability, i.e. we obtain the cis isomer to be more stable by $-0.24$\,eV than the trans configuration, which as such is an intriguing finding. It is again intuitively rationalized by the different preferred adsorption heights of the central azo-bridge and the closed-shell phenyl-rings, which can be much better acommodated in the three-dimensional cis form. For any strong enough azo-bridge to surface bonding this will naturally offset the higher gas-phase stability of the trans isomer, and as indicated by our calculations this seems to be the case at Cu(111). Nevertheless, the obtained preference by $-0.24$\,eV is a small number and has to be seen in the perspective that the GGA-PBE functional does not account for possibly different van der Waals contributions to the adsorption energy of the two isomers. This is particularly critical, as it is clear that these contributions will yield a larger stabilization for the more planar trans configuration, and would therewith counteract the covalent bond driven stability reversal. Our analysis of this issue on the level of the semi-empirical van der Waals correction schemes in fact indicates that the stability reversal does not prevail \cite{mcnellis09}. Nevertheless, in view of the unsure applicability of these schemes to the adsorption at metal surfaces it remains for future higher-level theory studies or experiment to fully settle this point. Not withstanding, despite the also partly significant changes to the geometric structure induced by the semi-empirical van der Waals correction schemes, we verified that this has only a minor effect on the electronic structure discussed in the next section. We correspondingly present this analysis at the optimized GGA-PBE geometries and only briefly comment on the changes obtained at the van der Waals corrected geometries.

\subsection{Electronic structure}

\begin{table}
\caption{\label{table4}
Electron affinity (EA), ionization potential (IP), and Mulliken electronegativity (EN) of the free trans and cis azobenzene molecule, compared to the work function of the clean surface ($\Phi$), the change of the work function induced by the adsorbed molecule ($\Delta \Phi$), and the change of the work function after subtracting the change induced by the potential drop through the molecular overlayer ($\Delta \Phi^*$), see text. The last two values correspond to the molecular coverage as resulting from the employed surface unit cell. Nevertheless, since we use the same $(6 \times 3)$ unit cell for both isomers, the induced work function changes are directly comparable. All values are given in eV.}
\begin{ruledtabular}
\begin{tabular}{ll|c|c}
          &                  &   trans azo    & cis azo   \\[0.1cm] \hline
EA        &                  &   1.06         & 0.75      \\
IP        &                  &   7.82         & 7.47      \\
EN        &                  &   4.44         & 4.11      \\[0.1cm] \hline
Cu(111)   &  $\Phi$          &   \multicolumn{2}{c}{4.83} \\
          &  $\Delta \Phi$   &   -0.69        & -0.96     \\
          &  $\Delta \Phi^*$ &   -0.15        & +0.11     \\[0.1cm] \hline
Ag(111)   &  $\Phi$          &   \multicolumn{2}{c}{4.46} \\
          &  $\Delta \Phi$   &   -0.26        & -0.77     \\
          &  $\Delta \Phi^*$ &   -0.25        & +0.03      \\[0.1cm] \hline
Au(111)   &  $\Phi$          &   \multicolumn{2}{c}{5.20} \\
          &  $\Delta \Phi$   &   -0.37        & -1.30     \\
          &  $\Delta \Phi^*$ &   -0.36        & -0.55     \\
\end{tabular}
\end{ruledtabular}
\end{table}

Important information concerning the bonding mechanism at the surface is already provided by the electronegativity (EN) of the isolated molecule. For trans and cis azobenzene we compute the Mulliken EN, defined as the mean of electron affinity and ionization potential, as 4.44\,eV and 4.11\,eV, respectively, cf. Table \ref{table4}. If we start our discussion again with the adsorption at Au(111) and correspondingly compare these values to the work function of the clean surface, $\Phi_{\rm Au(111)} = 5.20$\,eV, this suggests some charge transfer from the molecule to the metal substrate, and more specifically slightly more charge transfer in case of the less electronegative cis isomer. The actually calculated lowering of the work function for trans and cis azobenzene adsorption of $-0.37$\,eV and $-1.30$\,eV, respectively, is consistent with this expectation, cf. Table \ref{table4}. Quantitatively, it is, however, important to realize that even without any charge transfer (or more generally charge rearrangement) upon adsorption a work function change can already result if the molecular overlayer as such exhibits a non-zero dipole moment \cite{renzi05}. In this respect, the two azobenzene isomers differ significantly in that the gas-phase trans configuration has no dipole moment, whereas the cis configuration exhibits a finite calculated dipole moment of 3.23\,Debye. In order to disentangle the contributions to the work function change, we therefore compute the potential change across a free-standing molecular overlayer in exactly the same geometric structure as it has in the adsorbed state. Subtracting this term we arrive at the work function change $\Delta \Phi^*$ solely due to the charge rearrangement caused by the interaction of the overlayer with the metal substrate. As shown in Table \ref{table4}, the values for $\Delta \Phi^*_{\rm Au(111)}$ are $-0.36$\,eV and $-0.55$\,eV for trans and cis azobenzene, respectively, which now fit almost too perfectly into the electronegativity picture of an only slightly larger charge transfer to the metal in case of the cis isomer. At this stage, we stress, however, the well-known limitations of the charge transfer notion at metallic surfaces.\cite{scheffler00} Distinguishing between an actual transfer of charge between molecule and substrate or a polarization of the molecular charge accompanied by a varying degree of charge displacement due to Pauli repulsion in the overlapping electron density tails of the bonding partners is to some degree semantics.\cite{silva03} We make no attempt to contribute to this discussion here, and henceforth simply refer to charge transfer in quotes to emphasize its conceptual nature.

\begin{figure*}
  \centering
  \parbox{8cm}
  {
    \includegraphics[width=8cm]{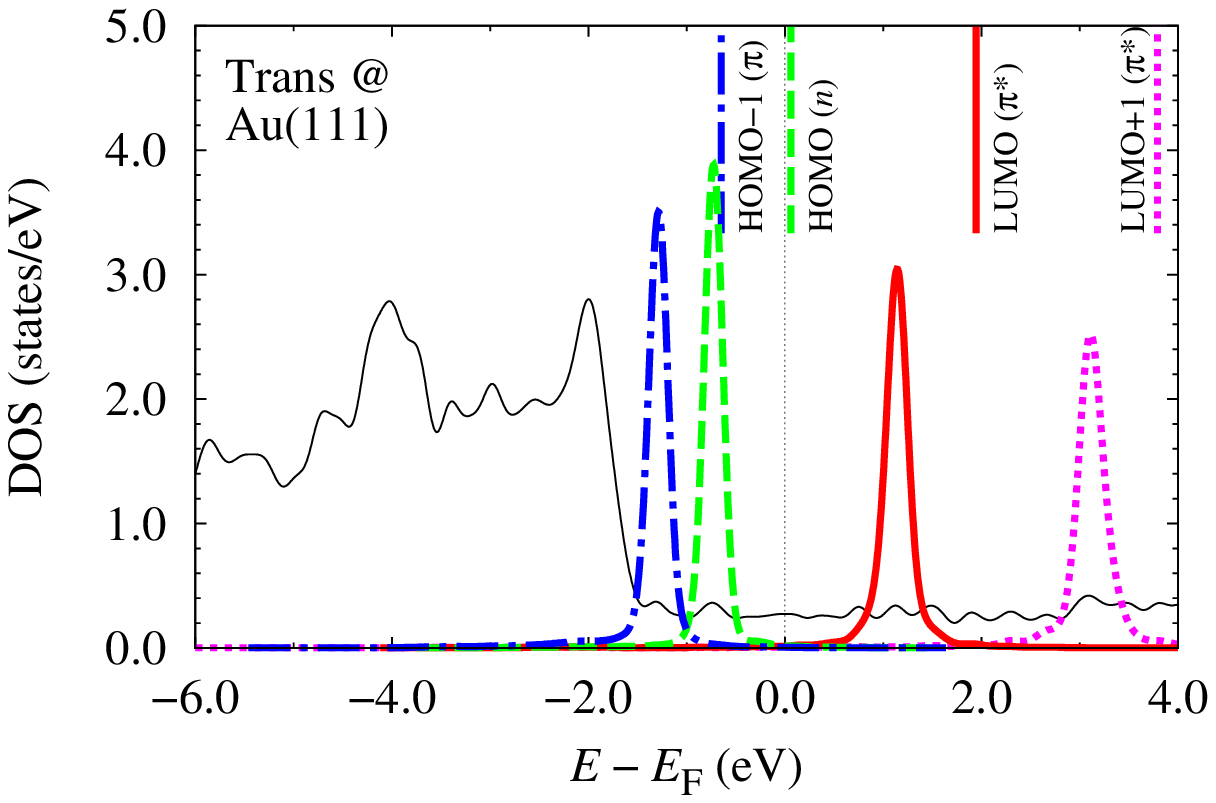}
  }
  \parbox{8cm}
  {
    \includegraphics[width=8cm]{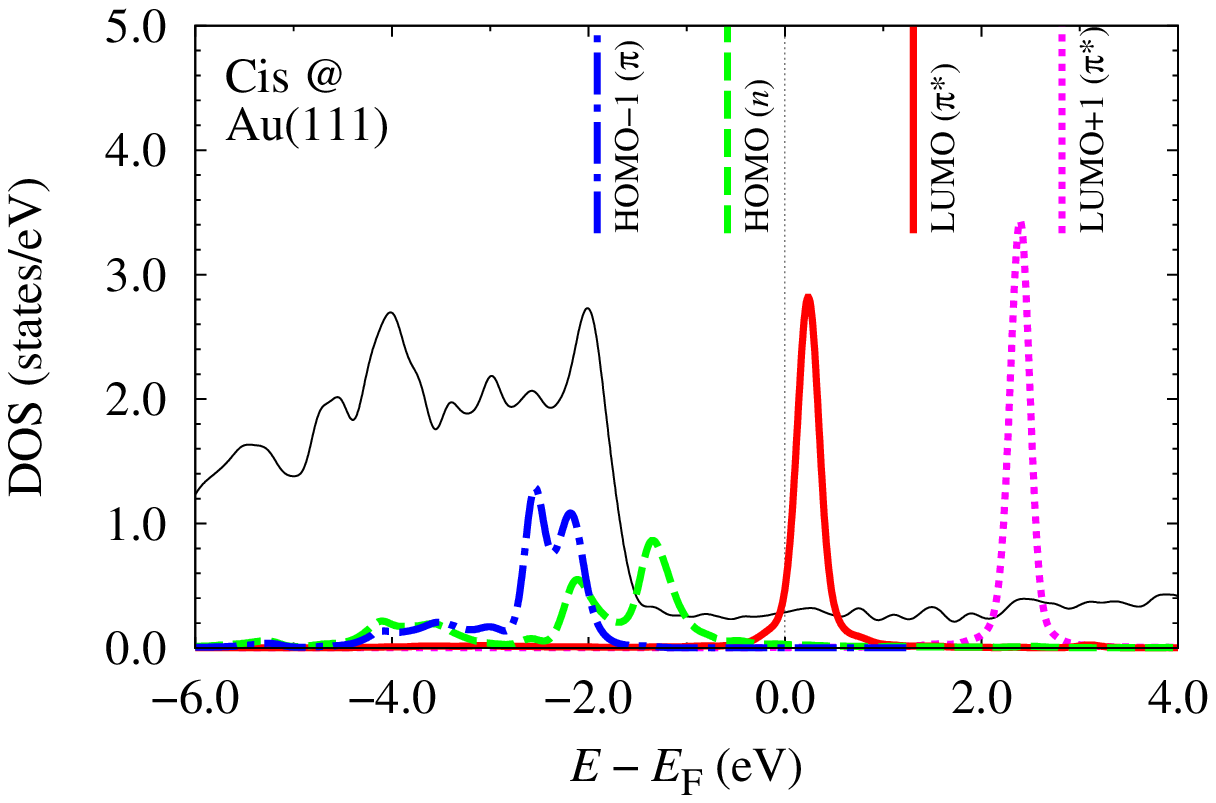}
  }
  \parbox{8cm}
  {
    \includegraphics[width=8cm]{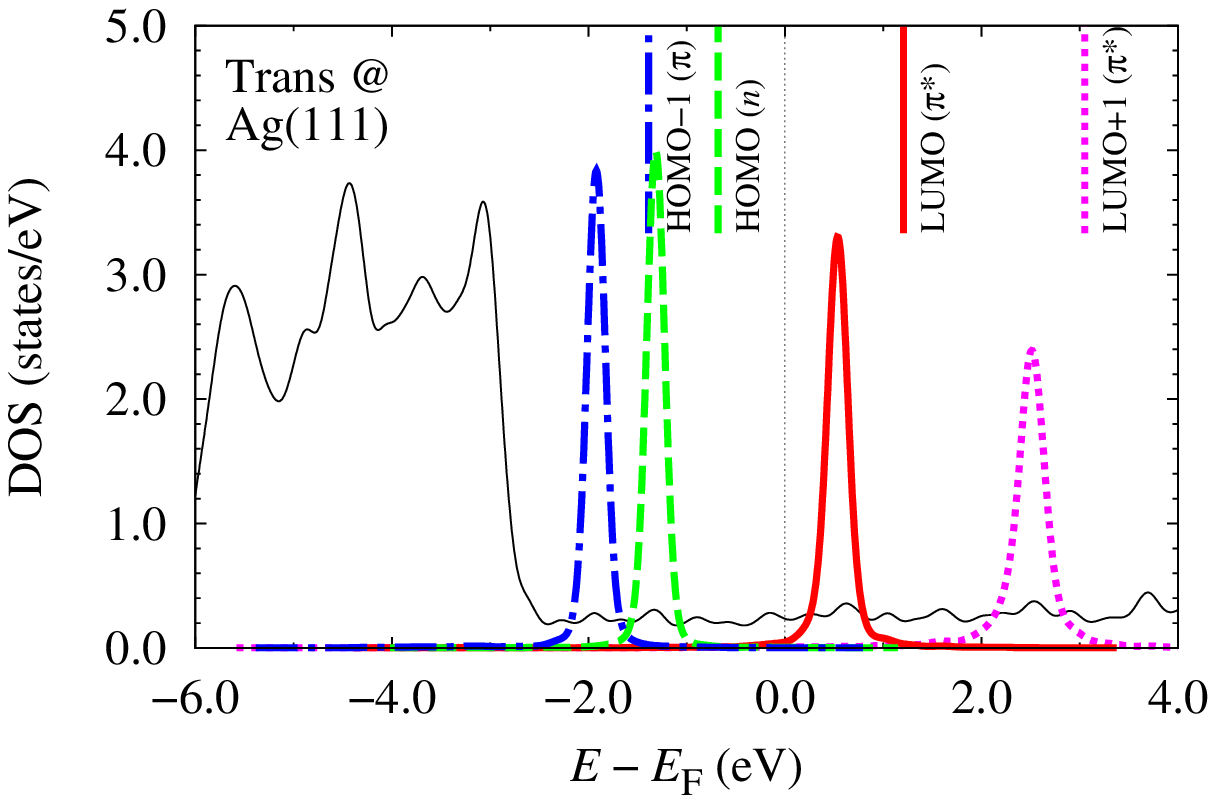}
  }
  \parbox{8cm}
  {
    \includegraphics[width=8cm]{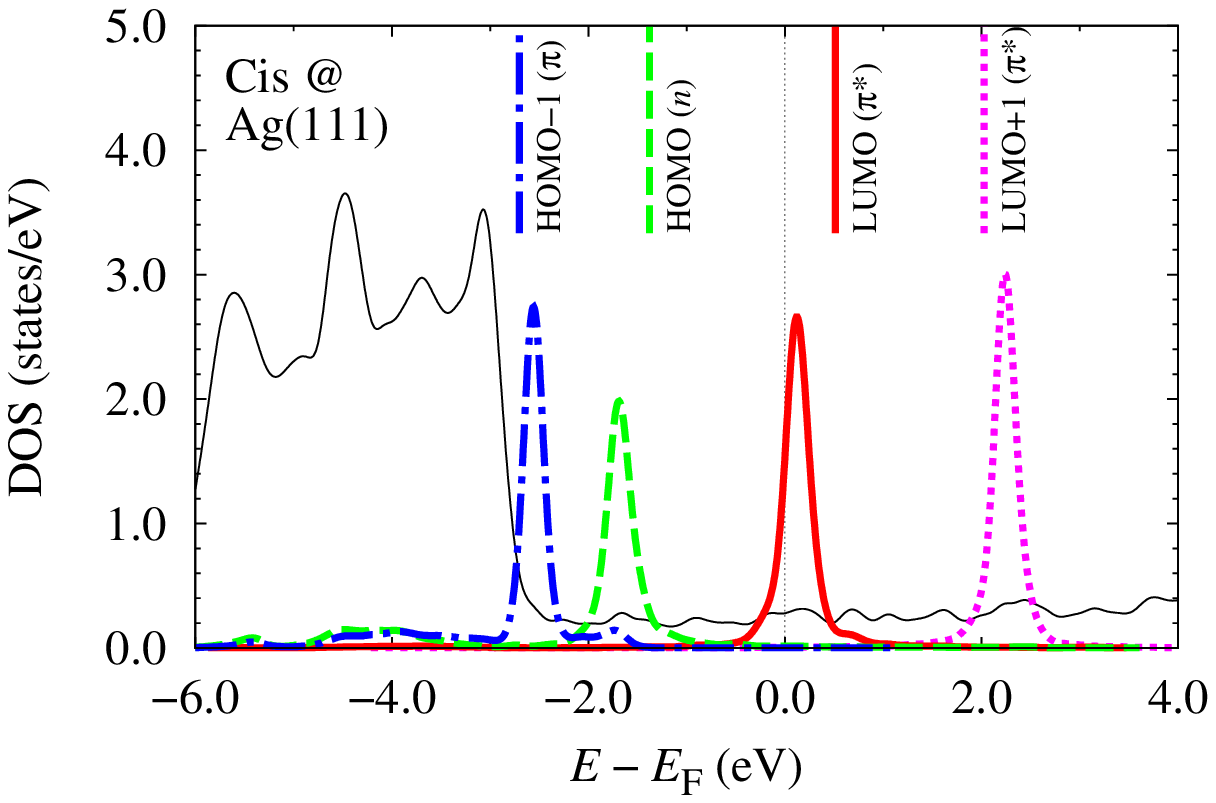}
  }
  \parbox{8cm}
  {
    \includegraphics[width=8cm]{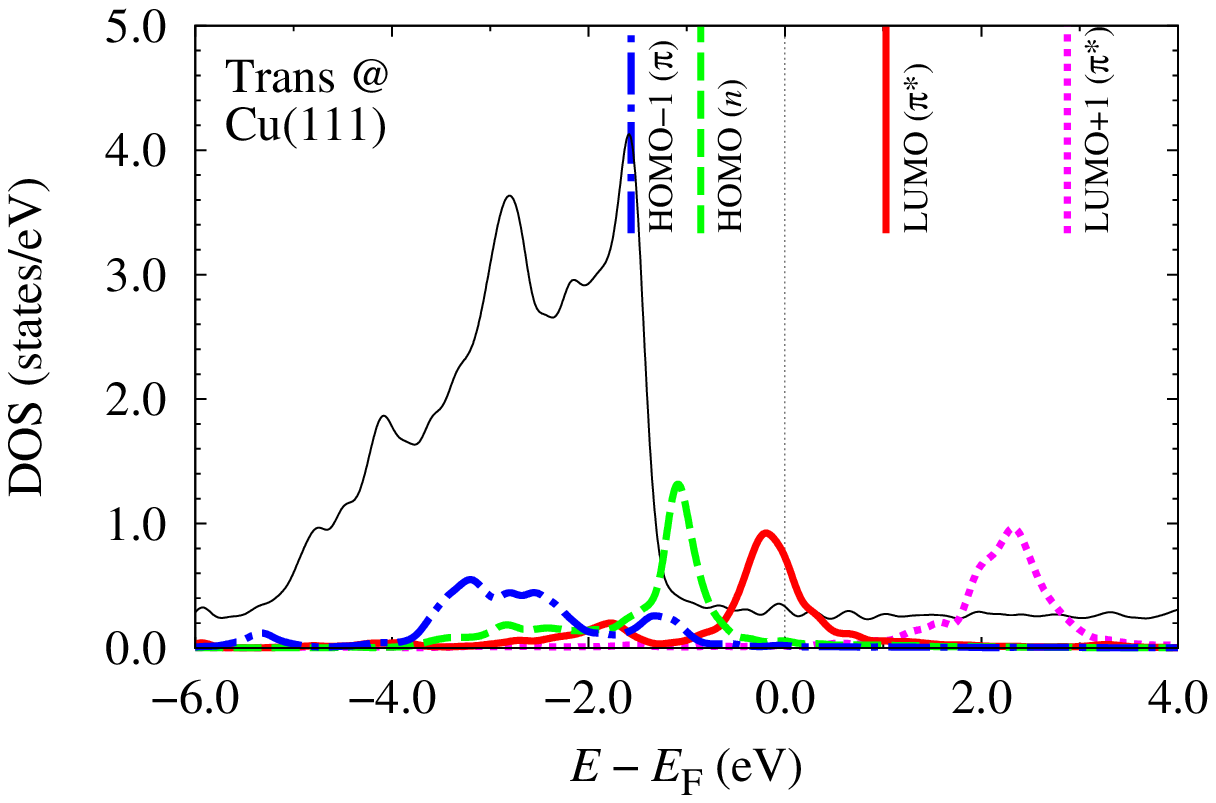}
  }
  \parbox{8cm}
  {
    \includegraphics[width=8cm]{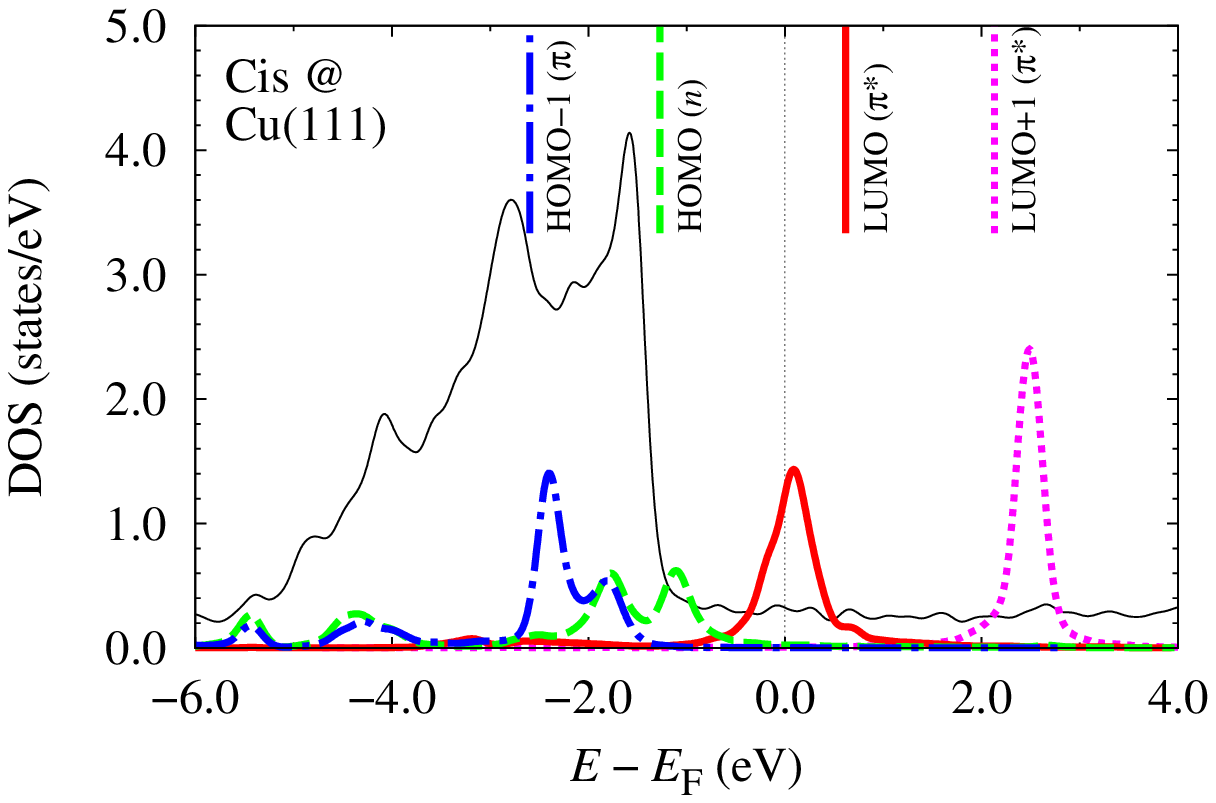}
  }
\caption{(Color online) Density of states (DOS) projected onto the frontier orbitals of gas-phase azobenzene: HOMO-1 (blue dotted line), HOMO (green dashed line), LUMO (red solid line), LUMO+1 (magenta dash-dotted line), see Fig. \ref{fig2} for a representation of these orbitals in the free molecule. Additionally shown is the total DOS of the employed supercell downscaled by a factor 75 (thin black line), as well as the position of the corresponding frontier orbitals (vertical sticks at the upper $x$-axis) in a modified Schottky-Mott picture that neglects the formation of a surface dipole through charge rearrangement upon adsorption (see text). Upper panel: Au(111). Middle panel: Ag(111), Lower panel: Cu(111). For all three surfaces adsorption of the trans isomer is shown on the left, and adsorption of the cis isomer on the right.}
\label{fig4}
\end{figure*}

Figure \ref{fig4} analyses the azobenzene/Au(111) interaction in terms of the density of states projected onto the molecular frontier orbitals. Similar to the discussion of the work function changes we aim to single out the effect of the charge rearrangement upon adsorption by also showing the level positions as they would result within a Schottky-Mott model, i.e. by simply aligning the vacuum level of the free molecule and the clean surface. While this model in general ignores the formation of a surface dipole, we modify it for our purposes by including the above described potential change caused already by the dipole moment of the free-standing molecular overlayer. Any shift between the orbital energy in this modified Schottky-Mott model and in the fully self-consistent calculation can this way be attributed to the charge rearrangement caused by the molecule-surface interaction. Starting with the PDOS of trans azobenzene at Au(111) shown in the left upper panel of Fig. \ref{fig4}, we see the four projected frontier orbitals, HOMO-1, HOMO, LUMO, and LUMO+1 broadened with hardly any degree of intermixing with Au $d$-states. Compared to their position in the modified Schottky-Mott model all levels are roughly equally shifted by about 0.8\,eV to lower energies, which is the direction one would also expect from the afore discussed ``charge transfer'' into the metal substrate, but which could equally well merely result from the lowering in the surface potential. Despite the down-shift, the LUMO still remains well above the Fermi-level. Both this and the actually resulting HOMO and LUMO positions at $-$0.7\,eV and +1.2\,eV respectively are in reasonable agreement with scanning-tunneling spectroscopy (STS) measurements by Choi {\em et al.} \cite{choi06}. This suggests that the sizable self-interaction induced error in the description of the HOMO-LUMO gap of the free molecule, cf. Section IIIA, is to some extent lifted, and that the orbital energies of the surface adsorbed molecule may be rather adequately described at the GGA-PBE level.

In case of cis azobenzene adsorption at Au(111) the actual level positions are in general at lower energies, cf. Fig. \ref{fig4}. This is primarily due to the fact that because of the finite potential change caused by the free-standing molecular overlayer already the level positions in the modified Schottky-Mott model lie at lower energies. With a comparable down-shift of the self-consistently calculated frontier orbitals as for trans azobenzene ({\em vide infra}), this has as one consequence that the projected LUMO lies now just above the Fermi-level. As a second consequence, both the HOMO and HOMO-1 level fall right into the energy range of the Au $d$-band with a concomitant stronger hybridization. The down-shift of the LUMO is with about 1.0\,eV even slightly stronger than for the trans isomer, again in line with the above discussed picture of a slightly stronger ``charge transfer'' as inferred from electronegativity differences and work function changes, or alternatively with a stronger lowering in the surface potential due to the lower adsorption height of this isomer. The exception of the much smaller shift of the LUMO+1 level of only $0.4$\,eV is in this respect primarily due to the distorted adsorption geometry: The computed LUMO+1 for a free molecule in this distorted geometry is higher in energy than in the relaxed gas-phase geometry, and if one takes this into account, one arrives at an effective down-shift of the LUMO+1 that is comparable to that of the other frontier orbitals.

Moving to the azobenzene adsorption at Ag(111) we find essentially the same picture as obtained at Au(111), yet with subtle and intriguing modifications. With a much lower work function of $\Phi_{\rm Ag(111)} = 4.46$\,eV the electronegativity difference to the azobenzene molecule is much smaller, and one would correspondingly expect a smaller ``charge transfer''. Consistently, the reduced work function change $\Delta \Phi^*$ upon trans azobenzene adsorption, i.e. after correcting for the dipole moment of the free-standing molecular overlayer, is with $-0.23$\,eV smaller than the corresponding value at Au(111), cf. Table \ref{table4}. Also the again roughly equal down-shift of the projected frontier orbitals compared to their position in the modified Schottky-Mott model shown in Fig. \ref{fig4} is with about 0.6\,eV smaller than the corresponding value of 0.8\,eV for trans adsorption at Au(111). Nevertheless, due to the much smaller Ag(111) work function the LUMO still ends up closer to the Fermi-level than was the case at Au(111), and is further lowered to just above $E_{\rm F}$ in the van der Waals corrected geometry. Nevertheless, with the Ag $d$-band starting at lower energies not even the HOMO-1 has significant overlap with the metal $d$-states, and we find all four frontier orbitals largely retaining their molecular character. While all of these findings fit perfectly into the overall picture developed at Au(111), the deviations start when turning to the cis azobenzene adsorption. From the smaller electronegativity of this isomer, one would expect a slightly larger ``charge transfer'' than in the case of trans azobenzene, and correspondingly a somewhat larger reduced work function change $\Delta \Phi^*$ as was the case at Au(111). However, actually calculated after correcting for the change induced by the free-standing molecular overlayer is an essentially zero $\Delta \Phi^*$, cf. Table \ref{table4}. We attribute this to the closeness of the LUMO to the Fermi-level already in the modified Schottky-Mott model as apparent in Fig. \ref{fig4}. A ``charge transfer'' or surface potential induced down-shift of this orbital leads therefore to a significant population of this formerly unoccupied level with a concomitant charge back-donation into the molecule counteracting the work function lowering.

This more complex interaction is finally also obtained for trans and cis azobenzene at Cu(111), where we now find a significant LUMO population for both isomers. That this results also for the trans case is partly already due to the strongly distorted adsorption geometries described in Section IIIB above. These distorted geometries exhibit a much increased dipole moment, which offsets the somewhat larger Cu(111) work function, cf. Table \ref{table3}, and leads to level positions in the modified Schottky-Mott model that are as low as in the case of Ag(111). With the higher-lying Cu $d$-band and the much shorter distance of the adsorbed molecule above the Cu(111) surface this enables a much stronger hybridization of the frontier orbitals than at the other two coinage metal surfaces, which is even further increased at the van der Waals corrected geometry. As a result HOMO-1, HOMO and LUMO all form broad bands with mixed metallic and molecular character and the counteracting donation and back-donation contributions result in overall small reduced work function changes $\Delta \Phi^*$ for both isomers, cf. Table \ref{table4}.

\section{Conclusions}

We have presented a DFT GGA-PBE trend study of the adsorption of azobenzene at the three coinage metal surfaces, Cu(111), Ag(111), and Au(111), providing detailed energetic, geometric, and electronic structure data of the meta-stable surface molecular states. As a central insight the calculations identify the conflicting bonding interests of the azo-bridge moiety on the one side and of the phenyl-rings on the other side as the governing factor for the molecule-surface interaction. While the phenyl-rings prefer a rather large, Pauli-repulsion dictated $z$-height above the surface, much shorter distances are required to enable the formation of stronger covalent N-metal bonds. For the planar trans configuration these two demands are incompatible and as a result we find the molecule either floating at a larger $z$-height above Au(111) and Ag(111), or severely buckled and distorted at a shorter $z$-height above Cu(111). The three-dimensional cis configuration does not suffer from this conflict and the actually calculated adsorption height corresponds at all three surfaces to N bond lengths to the directly coordinated surface metal atoms that are highly reminiscent of corresponding distances in transition metal complexes.

Due to the more favorable gas-phase molecular geometry adsorption leads thus naturally to a larger stabilization of the cis isomer. At the employed GGA-PBE level we correspondingly compute a diminished relative stability with the trans isomer still more stable at Au(111) and Ag(111), and an actual reversal of the gas-phase energetic order at Cu(111). This is traced back to the more reactive Cu $d$-band in conjunction with the smaller Cu atomic radius, which together enable the formation of strong covalent N-Cu bonds at a low adsorption height and a concomitantly reduced Pauli-repulsion of the phenyl-rings. From a critical discussion of these findings including a semi-empirical account of long-range van der Waals forces we conclude that the stronger dispersive attraction in case of the planar trans isomer largely counteracts the covalent bond driven preferential stabilization of the cis isomer. Whether the stability reversal of cis and trans azobenzene obtained with GGA-PBE at Cu(111) indeed prevails remains thus to be confirmed by higher-level theory or experiment.

Comparison to scarce experimental data suggests that the energetic positions of the frontier orbitals in the surface mounted geometry are reasonably well described to also attempt some conclusions with respect to the bonding and isomerization mechanisms. Light-induced switching in gas-phase and solution is conventionally discussed in terms of photo-excitation of the HOMO-1 to LUMO ($\pi \rightarrow \pi^*$) and HOMO to LUMO ($n \rightarrow \pi^*$) resonances in the trans and cis states, respectively.\cite{rau90} In this respect, the obtained noticeable broadening of these levels at all three surfaces casts severe doubts on the efficiency of these conventional mechanisms, and could rationalize the unsuccessful attempts to photo-isomerize bare azobenzene at Au(111) \cite{comstock07}. On the other hand, the actual degree of hybridization of the occupied orbitals could just be the key to the function, considering the HOMO-involving substrate-mediated charge-transfer process suggested recently in the context of the reversible photo-induced isomerization of the azobenzene-derivative TBA at Au(111) \cite{hagen08}. In this regard, we expect further insight from a detailed comparison of the present reference data for bare azobenzene with corresponding data for differently functionalized azobenzene molecules. A last intriguing aspect is the partial occupation of the LUMO in both isomers at Cu(111) and in the cis isomer at Ag(111) with a corresponding back-donation contribution to the bonding. It is thus only at Au(111) that the LUMO position remains distinctly above the metal Fermi-level for both isomers. This has quite some bearing on the negative ion resonance mechanism discussed in the context of tip-manipulation techniques \cite{fuechsel06} and could explain why the hitherto only such isomerization has been reported for this surface \cite{choi06}.

\section{Acknowledgements}

Funding by the Deutsche Forschungsgemeinschaft through Sfb 658 - Elementary Processes in Molecular Switches at Surfaces - is gratefully acknowledged. We thank the DEISA Consortium (co-funded by the EU, FP6 projects 508830/031513) for support within the DEISA Extreme Computing Initiative (www.deisa.org).

\end{document}